\documentclass[amsmath,12pt,amssymb,preprint,nofootinbib]{revtex4}
\usepackage{amsfonts} %\usepackage{citesort}
\usepackage{graphicx} % Include figure files
\usepackage{epsfig}
\usepackage{multirow}
\usepackage{bm}% bold math

\begin{document}

\title{\textbf
{Masses of Heavy Quarkonium states in magnetized matter \\
- effects of PV mixing and (inverse) magnetic catalysis}}
\author{Ankit Kumar}
\email{ankitchahal17795@gmail.com} 
\author{Amruta Mishra}
\email{amruta@physics.iitd.ac.in}  
\affiliation{Department of Physics, 
Indian Institute of Technology, Delhi, New Delhi - 110016}

\begin{abstract}
We study the in-medium masses of the heavy quarkonium (charmonium and 
bottomonium) states in isospin asymmetric nuclear matter 
in presence of an external magnetic field.
The mass modifications of the heavy quarkonia are obtained
from the medium modifications of a scalar dilaton field, $\chi$,
calculated within a chiral effective model. The dilaton field is
introduced in the model through a scale invariance breaking 
logarithmic potential, and, simulates the gluon condensates of QCD.
Within the chiral effective model, the values of the dilaton field
along with the scalar (isoscalar, $\sigma (\sim \langle \bar u u \rangle
+\langle \bar d d \rangle)$, isoscalar $\zeta (\sim \langle \bar s s \rangle$)
and isovector $\delta (\sim \langle  \bar u u\rangle
 - \langle\bar d d \rangle)$)
fields, are solved from their coupled  equations of motion.
These are solved accounting for the effects of the Dirac sea (DS)
as well as anomalous magnetic moments (AMMs) of the nucleons. 
When AMMs are negelcted, both at zero density and at nuclear matter
saturation density, $\rho_0$, the Dirac sea contributions are observed 
to lead to enhancement of the quark condensates (through $\sigma$ and
$\zeta$ fields) with increase in magnetic field, 
an effect called the magnetic catalysis (MC). However, the inclusion
of AMMS is observed to lead to the opposite trend of
inverse magnetic catalysis (IMC) for $\rho_B=\rho_0$.
The magnetic field effects on the masses
of the heavy quarkonia include the mixing of the pseudoscalar 
(spin 0) and vector (spin 1) states (PV mixing), 
as well as, the effects from (inverse) magnetic catalysis.
These effects are observed to be significant 
for large values of the magnetic field. This should 
have observable consequences on the production 
of the heavy quarkonia and open heavy flavour mesons,
resulting from ultra-relativistic peripheral heavy ion collision 
experiments, where the created magnetic field can be huge. 
\end{abstract}
\maketitle
\vspace{-1cm}
\section{Introduction}
The study of the in-medium properties of the heavy quarkonia
is a topic of intense research work
due to its relevance in heavy ion collision experiments
\cite{Hosaka}. The magnetic fields created in peripheral 
ultra-relativistic heavy ion collision experiments, e.g.,
at LHC, CERN and RHIC, BNL are estimated to be huge \cite{tuchin}.
This has initiated a lot of work in the study of hadrons, 
in particular, of the heavy quarkonia as well as open heavy 
flavour mesons in the presence of magnetic fields, 
due to the reason that these heavy mesons are created 
at the early stage when the magnetic field can still be large.
The heavy quarkonium (charmonium and bottomonium) states 
have been investigated in the literature 
using the potential models 
\cite{eichten_1,eichten_2,satz_1,satz_2,satz_3,satz_4,satz_5,repko,
Ebert,Bonati_pot_model,Yoshida_Suzuki_heavy_flavour_meson_strong_B},
the QCD sum rule approach
\cite{kimlee,klingl,amarvjpsi_qsr,jpsi_etac_mag,moritalee_1,moritalee_2,moritalee_3,moritalee_4,open_heavy_flavour_qsr_1,open_heavy_flavour_qsr_2,open_heavy_flavour_qsr_3,open_heavy_flavour_qsr_4,Wang_heavy_mesons_1,Wang_heavy_mesons_2,arvind_heavy_mesons_QSR_1,arvind_heavy_mesons_QSR_2,arvind_heavy_mesons_QSR_3},
the coupled channel approach
\cite{ltolos,ljhs,mizutani_1,mizutani_2,HL,tolos_heavy_mesons_1,tolos_heavy_mesons_2}, the quark meson coupling (QMC) model
\cite{open_heavy_flavour_qmc_1,open_heavy_flavour_qmc_2,open_heavy_flavour_qmc_3,qmc_1,qmc_2,qmc_3,qmc_4,krein_jpsi,krein_17},
heavy quark symmetry and interaction
of these mesons with nucleons via pion exchange \cite{Yasui_Sudoh_pion},
heavy meson effective theory
\cite{Yasui_Sudoh_heavy_meson_Eff_th}, studying the heavy flavour meson as
an impurity in nuclear matter \cite{Yasui_Sudoh_heavy_particle_impurity}.

Using leading order QCD formula \cite{pes1,pes2,voloshin},
the mass modifications of the charmonium states
were calculated in a linear density approximation in Ref.\cite{leeko},
due to the medium change of the scalar gluon condensate.
The study showed much larger mass shifts for the excited states,
$\psi(2S)$ and $\psi(1D)$, as compared to the mass shift of $J/\psi$.
Within a chiral effective model \cite{Schechter,paper3,kristof1},
generalized to include the interactions of the charm and bottom
flavoured hadrons,
the in-medium heavy quarkonium (charmonium and bottomonium) masses are
obtained from the medium changes of a scalar dilaton field,
which mimics the gluon condensates of QCD
\cite{amarvdmesonTprc,amarvepja,AM_DP_upsilon}.
The mass modifications of  the open heavy flavour
(charm and bottom) mesons within the chiral effective model
have also been studied from their interactions with the baryons
and scalar mesons in the hadronic medium
\cite{amdmeson,amarindamprc,amarvdmesonTprc,amarvepja,DP_AM_Ds,DP_AM_bbar,DP_AM_Bs}.
The chiral effective model, in the original version
with three flavours of quarks (SU(3) model),
 has been used extensively in the literature,
for the study of finite nuclei \cite{paper3},
strange hadronic matter \cite{kristof1},
light vector mesons \cite{hartree},
strange pseudoscalar mesons, e.g. the kaons and antikaons
\cite{kaon_antikaon,isoamss,isoamss1,isoamss2}
in isospin asymmetric hadronic matter,
as well as for the study of bulk matter of neutron stars
\cite{pneutronstar}.
Using the medium changes of the light iuuark condensates
and gluon condensates calculated within the
chiral SU(3) model,
the light vector mesons ($\omega$, $\rho$ and $\phi$)
in (magnetized) hadronic matter have been
studied within the framework of QCD sum rule approach
 \cite{am_vecmeson_qsr,vecqsr_mag}.
The kaons and antikaons have been recently studied
in the presence of strong magnetic fields
using this model \cite{kmeson_mag}.
The model has been used to study the partial decay widths
of the heavy quarkonium states to the open heavy flavour mesons,
in the hadronic medium
\cite{amarvepja} using a light quark creation model \cite{friman},
namely the $^3P_0$ model \cite{3p0_1,3p0_2,3p0_3,3p0_4} as well as
using a field theoretical model for composite hadrons
\cite{amspmwg,amspm_upsilon}.
Recently, the effects of magnetic field on the charmonium
partial decay widths to $D\bar D$ mesons have been
studied using the $^3P_0$ model \cite{charmdecay_mag}
and, charmonium (bottomonium) decay widths to $D\bar D$ 
($B\bar B$) using the field
theoretic model of composite hadrons 
\cite{charmdw_mag,upslndw_mag}.

In the present work, we study the modifications of the masses 
of the charmonium ($J/\psi$, $\psi(2S)$ and $\psi(1D)$)
and the bottomonium ($\Upsilon(1S)$, $\Upsilon(2S)$, $\Upsilon(3S)$
and $\Upsilon(4S)$) states in magnetized (nuclear) matter,
within a chiral effective model. The mass modifications 
of these states arise due to the medium change of a dilaton field,
which is incorporated into the model to simulate the gluon condensates
of QCD. Using the approximation that the scalar fields
%($\sigma \sim \langle \bar u u +\bar d d \rangle$,
%$\zeta \sim \langle \bar s s \rangle$ and
%$\delta \sim \langle \bar u u -\bar d d \rangle$),
are treated as classical fields, the value of the dilaton field,
$\chi$, is obtained from solution of the coupled equations of motion
of the scalar (nonstrange isoscalr $\sigma$,
strange isoscalar $\zeta$ and nonstrange isovector $\delta$) 
fields and $\chi$. 
The Dirac sea contributions of the nucleons are also taken into 
consideration for obtaining the nucleon self-energy, 
by summing over the tadpole diagrams. The effects of the
anomalous magnetic moments (AMMs) of the nucleons
are observed to lead to important modifications
to the self energy of the nucleons.  
Within the chiral effective model, for zero density
as well as for the baryon density, $\rho_B=\rho_0$, the nuclear matter
saturation density, when the AMMs are not taken into account,
both for symmetric and asymmetric nuclear matter, 
the Dirac sea contributions
are observed to lead to an enhancement of the quark condensates 
(through the scalar $\sigma \sim (\langle  \bar u u +\bar d d \rangle)$ 
and $\zeta (\sim \langle \bar s s\rangle )$ fields), 
with increase in the magnetic field,
an effect called `magnetic catalysis (MC)',
However, with inclusion of the AMMS of the nucleons,
for $\rho_B=\rho_0$, the opposite trend, i.e., the inverse magnetic
cataalysis (IMC) is observed within the chiral model.
The increase of the light quark condensates with magnetic field, 
has been studied in a large extent on the quark matter sector 
using the Nambu-Jona-Lasinio (NJL) model \cite{Preis,menezes}. 
In Ref. \cite{haber}, the effect of magnetic catalysis has been 
studied for the nuclear matter using the Walecka model 
and an extended linear sigma model.
In Ref. \cite{arghya}, the effects of magnetic field have been 
studied in the Walecka model by using a weak field approximation 
of the fermion propagator. The effect of the anomalous magnetic 
moment of the nucleons are seen to enhance the catalysis effect 
at zero temperature and zero baryon density
\cite{arghya}. However, at finite temperature, the critical 
temperature for the vacuum to nuclear matter phase transition 
for nonzero anomalous magnetic moment of the nucleons, is seen 
to rise with increasing magnetic field, implying inverse 
magnetic catalysis \cite{balicm}, whereas for vanishing
AMMs of the nucleons, the behavior is opposite, 
indicating the magnetic catalysis. 
Thus, the effect of the anomalous magnetic moments of the 
nucleons are important to study the contributions
from the Dirac sea in presence of finite magnetic field. 
In the literature there are very few works on the magnetic 
catalysis effect in the nuclear matter. In the present work, 
we have incorporated the effects of the Dirac sea through 
summation of nucleonic tadpole diagrams within a chiral
effective model. In the present study of the heavy
quarkonia masses, we also consider the mixing of the pseudoscalar and 
vector meson (PV mixing)
\cite{charmdw_mag,upslndw_mag,open_charm_mag_AM_SPM,strange_AM_SPM}.

The outline of the paper is as follows.
In section II, we describe breifly the computaiton of the 
mass modificaitons of the heavy quarkonium states 
in magnetized (nuclear) matter using a chiral effective model.
These masses are calculated within the model from the modification 
of the scalar dilaton, which mimics the scale symmetry breaking of QCD.
The magnetic field effects considered are the pseudoscalar - vector 
meson (PV) mixing and the contributions of the 
Dirac sea of the nucleons. The PV mixing
corresponds to the mixing of the pseudoscalar (spin 0) and the vector
(spin 1) mesons in the presence of a magnetic field.   
The latter leads to the (inverse) magentic catalysis effect.
In section III, the results of the medium modifications of the 
charmonium and bottomonium masses in magnetized matter
are discussed and section IV summarizes the findings of this work.

\section{Mass modifications of heavy Quarkonium states in magnetized matter}

The in-medium masses of the charmonium ($J/\psi$, $\psi(2S)$
and $\psi(1D)$) and bottomonium states ($\Upsilon(1S)$, $\Upsilon(2S)$, 
$\Upsilon(3S)$ and $\Upsilon(4S)$) are studied in magnetized (nuclear)
matter. The effects of pseudoscalar - vector meson (PV) mixing 
($J/\psi-\eta_c$, $\psi(2S)-\eta_c (2S)$, $\psi(1D)-\eta_c (2S)$
for charmonium states and $\Upsilon (1S)-\eta_b$, 
$\Upsilon (2S)-\eta_b(2S)$, $\Upsilon (3S)-\eta_b(3S)$,
$\Upsilon (4S)-\eta_b(4S)$ for the bottomonium states),
and the Dirac sea contributions for the nucleons
are considered in the present study 
of mass modifications of these heavy mesons in the presence 
of a magnetic field.

\begin{figure}
\vskip -3.2in
    \includegraphics[width=1.\textwidth]{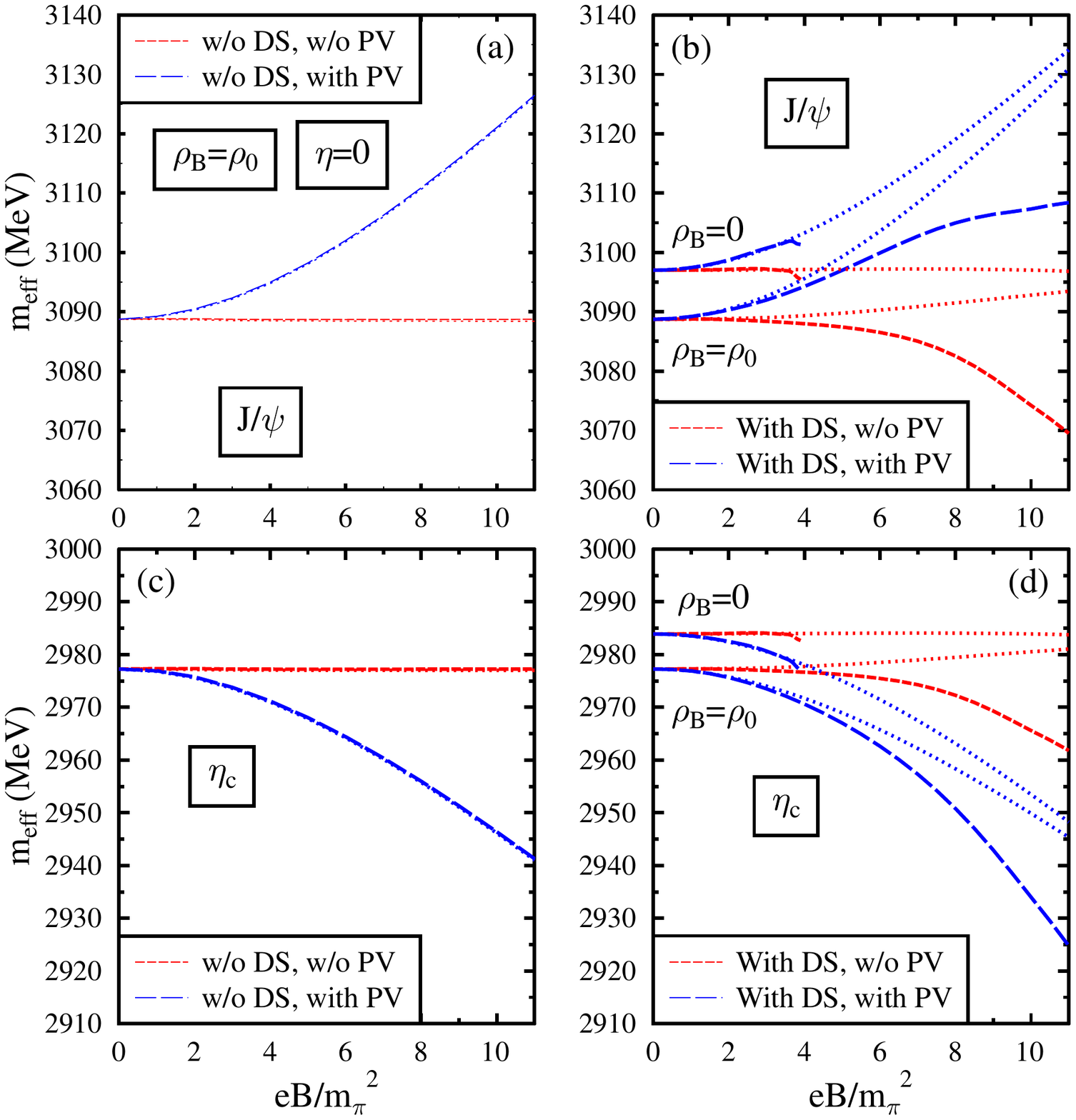}\hfill
\vskip -0.3in
    \caption{Masses (MeV) of $J/\psi$ and $\eta_c$ are plotted as
functions of $eB/m_\pi^2$ for $\rho_B=\rho_0$ and $\eta=0$,
with and without the effects of PV mixing. 
These are shown in (a) and (c), when the Dirac sea contributions 
of nucleons (resulting in (inverse) magnetic catalysis)
are not included and in (b) and (d), while these
effects are included. In (b) and (c), the masses are also shown
for $\rho_B=0$ due to the Dirac sea conttributions.
The results are for cases when the AMMs are included,
which are compared with the cases when AMMs are not considered
(shown as dotted lines).}
    \label{mjpsi_etac_rhb0_eta0_MC}
\end{figure}

\begin{figure}
\vskip -3.2in
    \includegraphics[width=1.\textwidth]{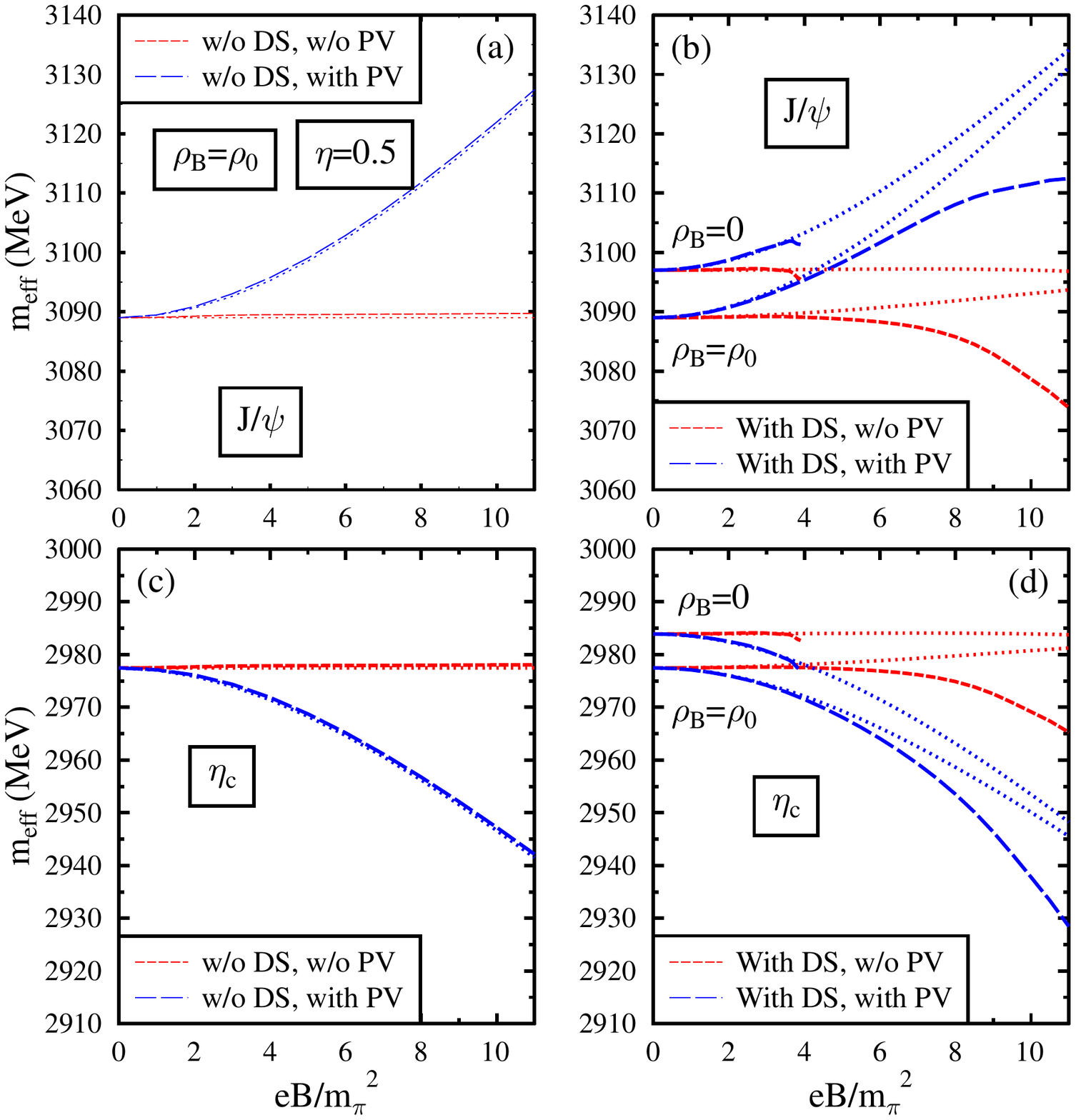}\hfill
\vskip -0.3in
    \caption{Same as Figure \ref{mjpsi_etac_rhb0_eta0_MC}, but with
$\eta$=0.5.}
    \label{mjpsi_etac_rhb0_eta5_MC}
\end{figure}

\begin{figure}
\vskip -3.2in
    \includegraphics[width=1.\textwidth]{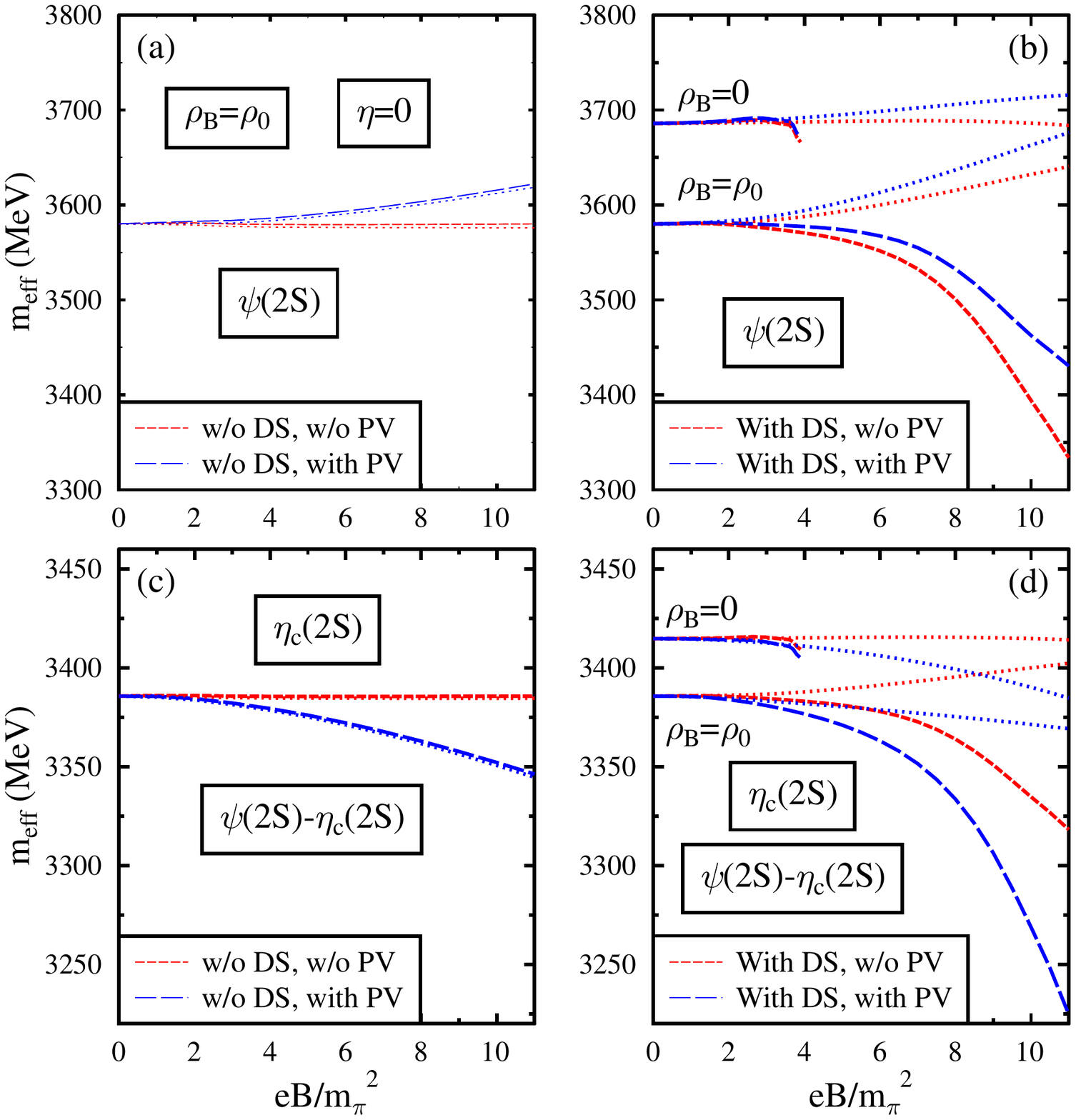}\hfill
\vskip -0.3in
    \caption{Masses (MeV) of $\psi(2S)$ and $\eta_c(2S)$ are plotted as
functions of $eB/m_\pi^2$ for $\rho_B=\rho_0$ and $\eta=0$,
with and without the effects of PV mixing. 
These are shown in (a) and (c), when the Dirac sea contributions 
of nucleons (resulting in  (inverse) magnetic catalysis)
are not included and in (b) and (d), while these
effects are included. In (b) and (c), the masses are also shown
for $\rho_B=0$ due to the Dirac sea conttributions.
The results are for cases when the AMMs are included,
which are compared with the cases when AMMs are not considered
(shown as dotted lines).}
    \label{mpsip_etacp_rhb0_eta0_MC}
\end{figure}

\begin{figure}
\vskip -3.2in
    \includegraphics[width=1.\textwidth]{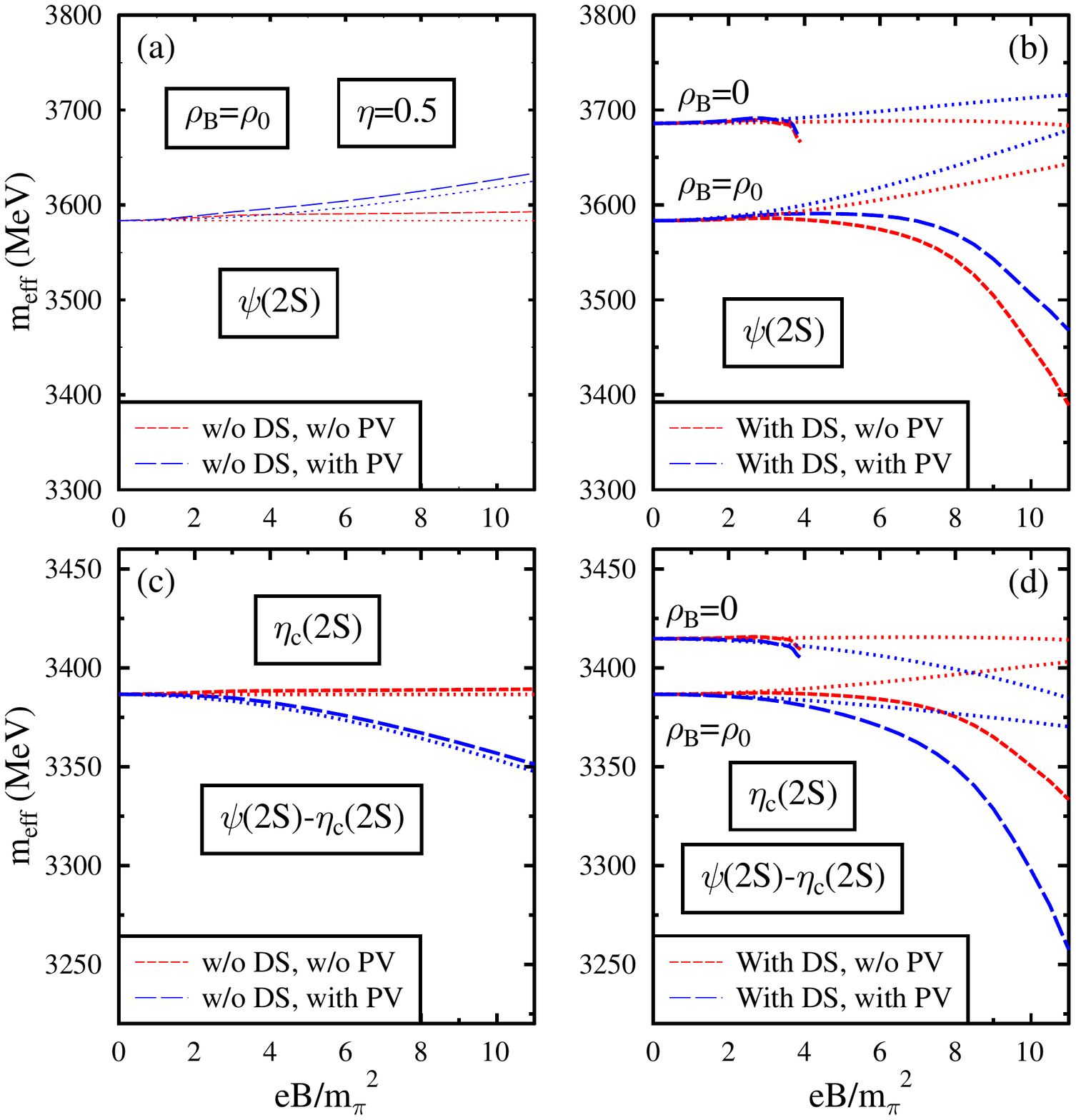}\hfill
\vskip -0.3in
    \caption{Same as Figure \ref{mpsip_etacp_rhb0_eta0_MC}, but with
$\eta$=0.5.}
    \label{mpsip_etacp_rhb0_eta5_MC}
\end{figure}

\begin{figure}
\vskip -3.2in
    \includegraphics[width=1.\textwidth]{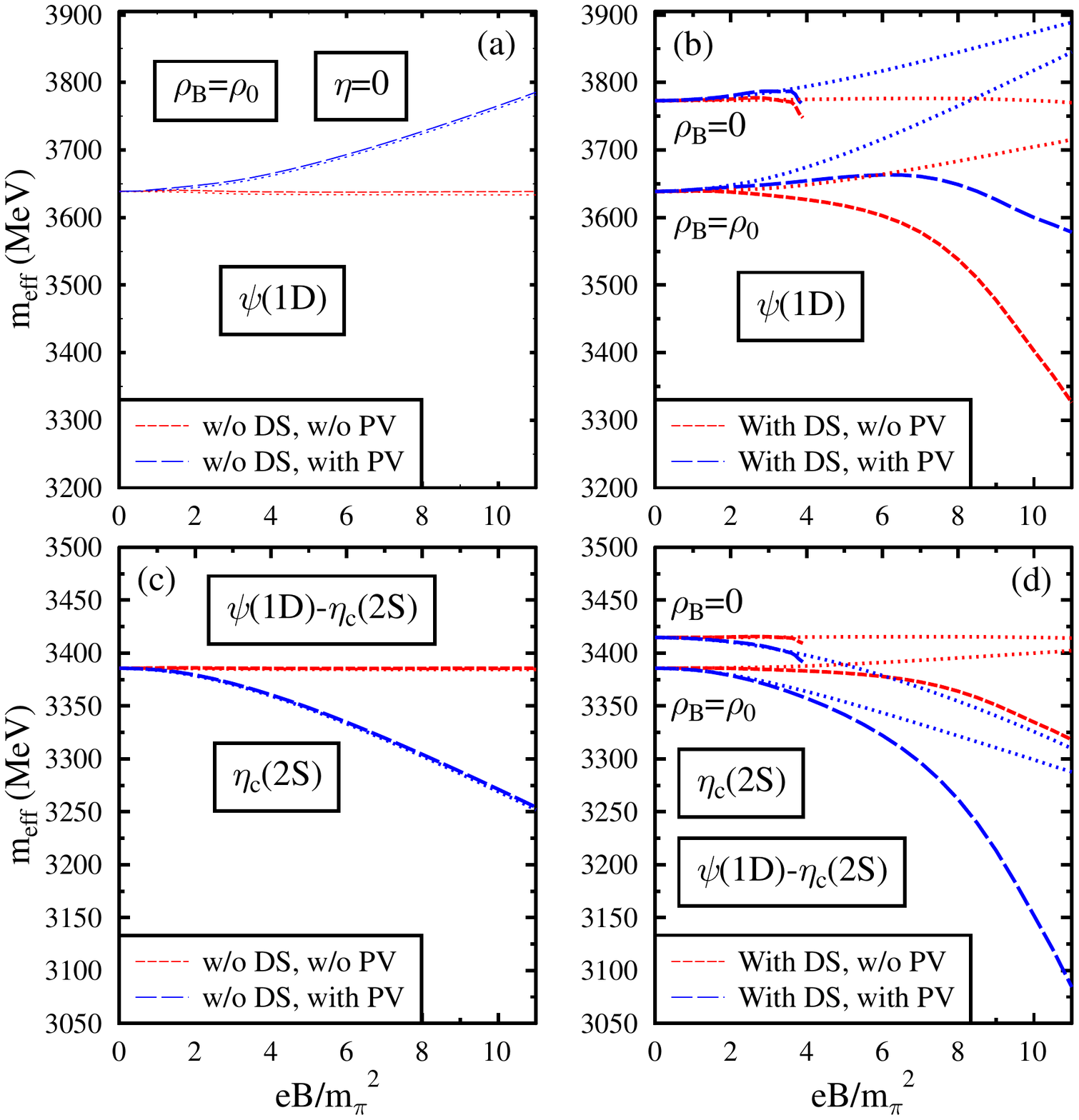}\hfill
\vskip -0.3in
    \caption{Masses (MeV) of $\psi(1D)$ and $\eta_c(2S)$ are plotted as
functions of $eB/m_\pi^2$ for $\rho_B=\rho_0$ and $\eta=0$,
with and without the effects of PV mixing. 
These are shown in (a) and (c), when the Dirac sea contributions 
of nucleons (resulting in (inverse) magnetic catalysis)
are not included and in (b) and (d), while these
effects are included. In (b) and (c), the masses are also shown
for $\rho_B=0$ due to the Dirac sea conttributions.
The results are for cases when the AMMs are included,
which are compared with the cases when AMMs are not considered
(shown as dotted lines).}
    \label{mpsi1d_etacp1d_rhb0_eta0_MC}
\end{figure}

\begin{figure}
\vskip -3.2in
    \includegraphics[width=1.\textwidth]{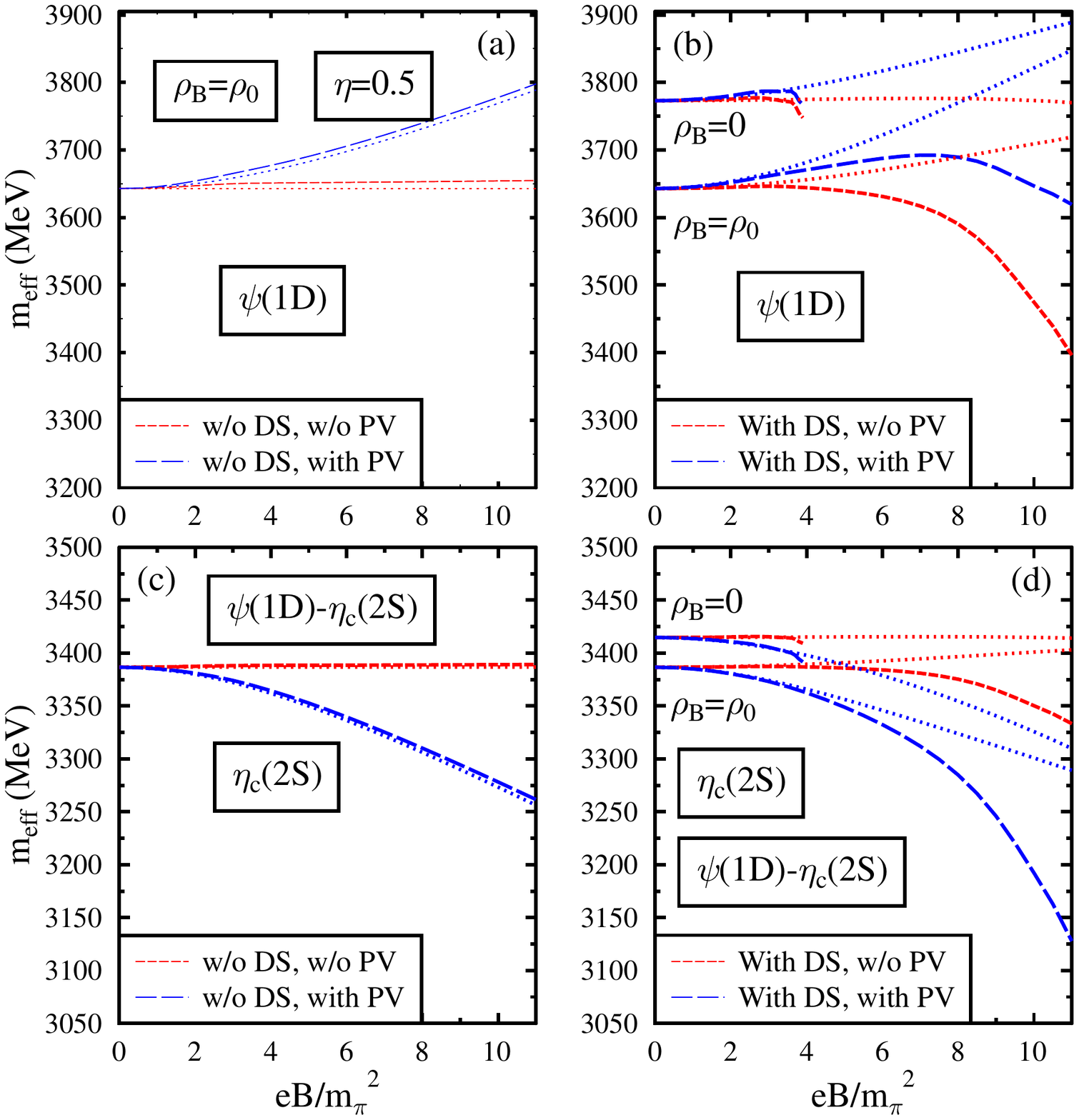}\hfill
\vskip -0.3in
    \caption{Same as Figure \ref{mpsi1d_etacp1d_rhb0_eta0_MC}, but with
$\eta$=0.5.}
    \label{mpsi1d_etacp1d_rhb0_eta5_MC}
\end{figure}

\begin{figure}
\vskip -3.2in
    \includegraphics[width=1.\textwidth]{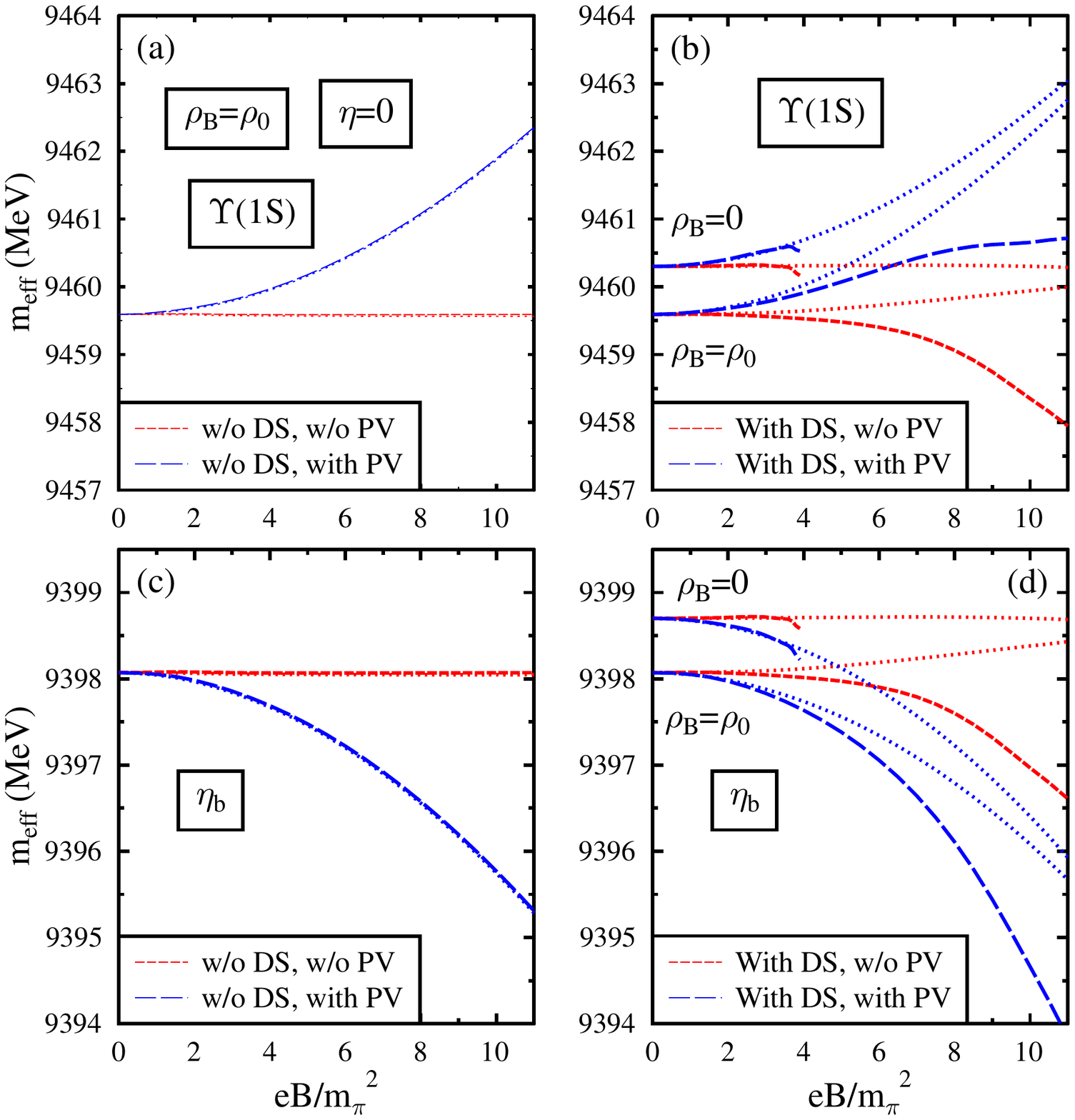}\hfill
\vskip -0.3in
    \caption{Masses (MeV) of $\Upsilon (1S)$ and $\eta_b$ are plotted as
functions of $eB/m_\pi^2$ for $\rho_B=\rho_0$ and $\eta=0$,
with and without the effects of PV mixing. 
These are shown in (a) and (c), when the Dirac sea contributions 
of nucleons (resulting in  (inverse) magnetic catalysis)
are not included and in (b) and (d), while these
effects are included. In (b) and (c), the masses are also shown
for $\rho_B=0$ due to the Dirac sea conttributions.
The results are for cases when the AMMs are included,
which are compared with the cases when AMMs are not considered
(shown as dotted lines).}
    \label{mups1s_etab_rhb0_eta0_MC}
\end{figure}

\begin{figure}
\vskip -3.2in
    \includegraphics[width=1.\textwidth]{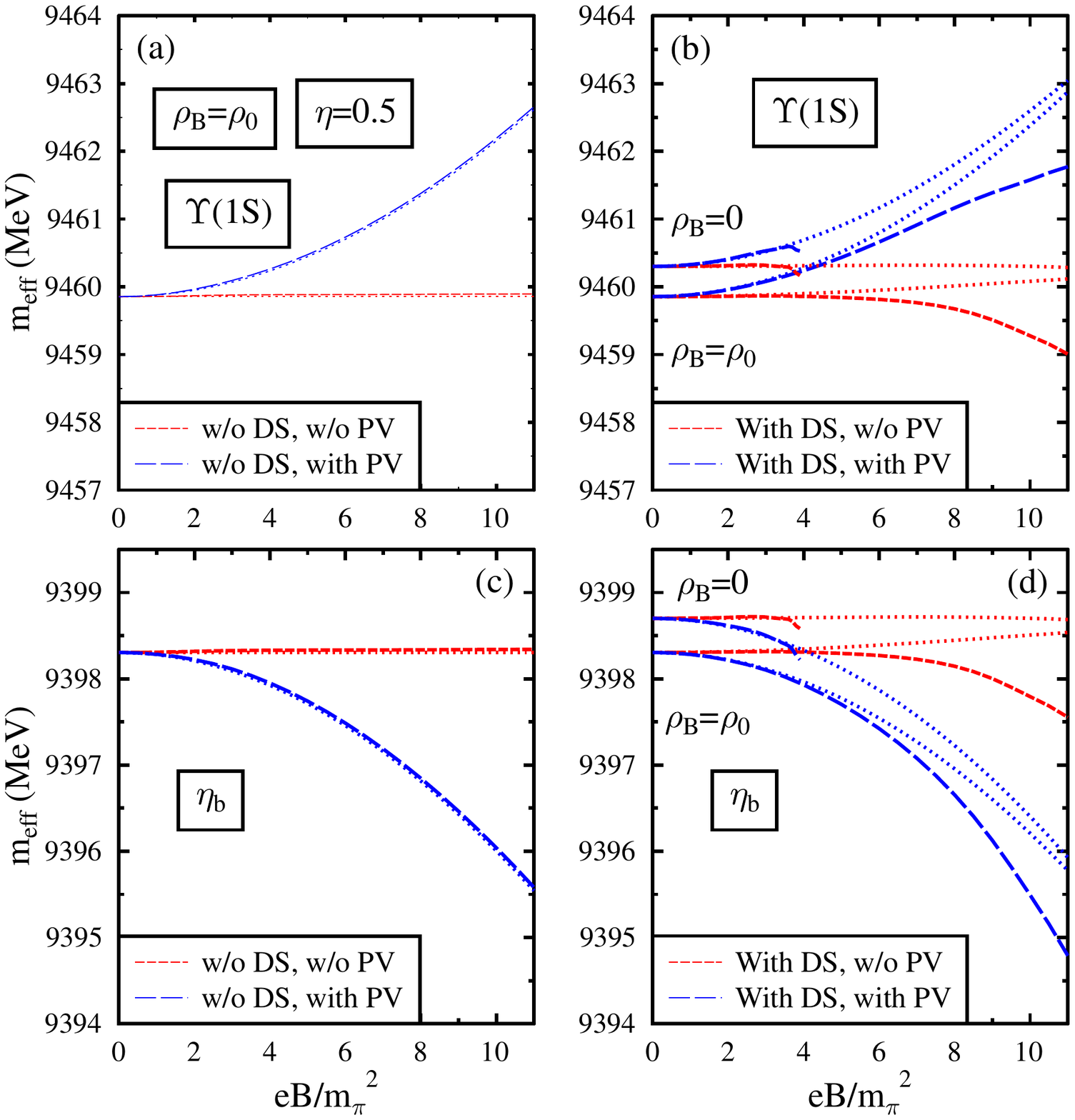}\hfill
\vskip -0.3in
    \caption{Same as Figure \ref{mups1s_etab_rhb0_eta0_MC}, but with
$\eta$=0.5.}
    \label{mups1s_etab_rhb0_eta5_MC}
\end{figure}

\begin{figure}
\vskip -3.2in
    \includegraphics[width=1.\textwidth]{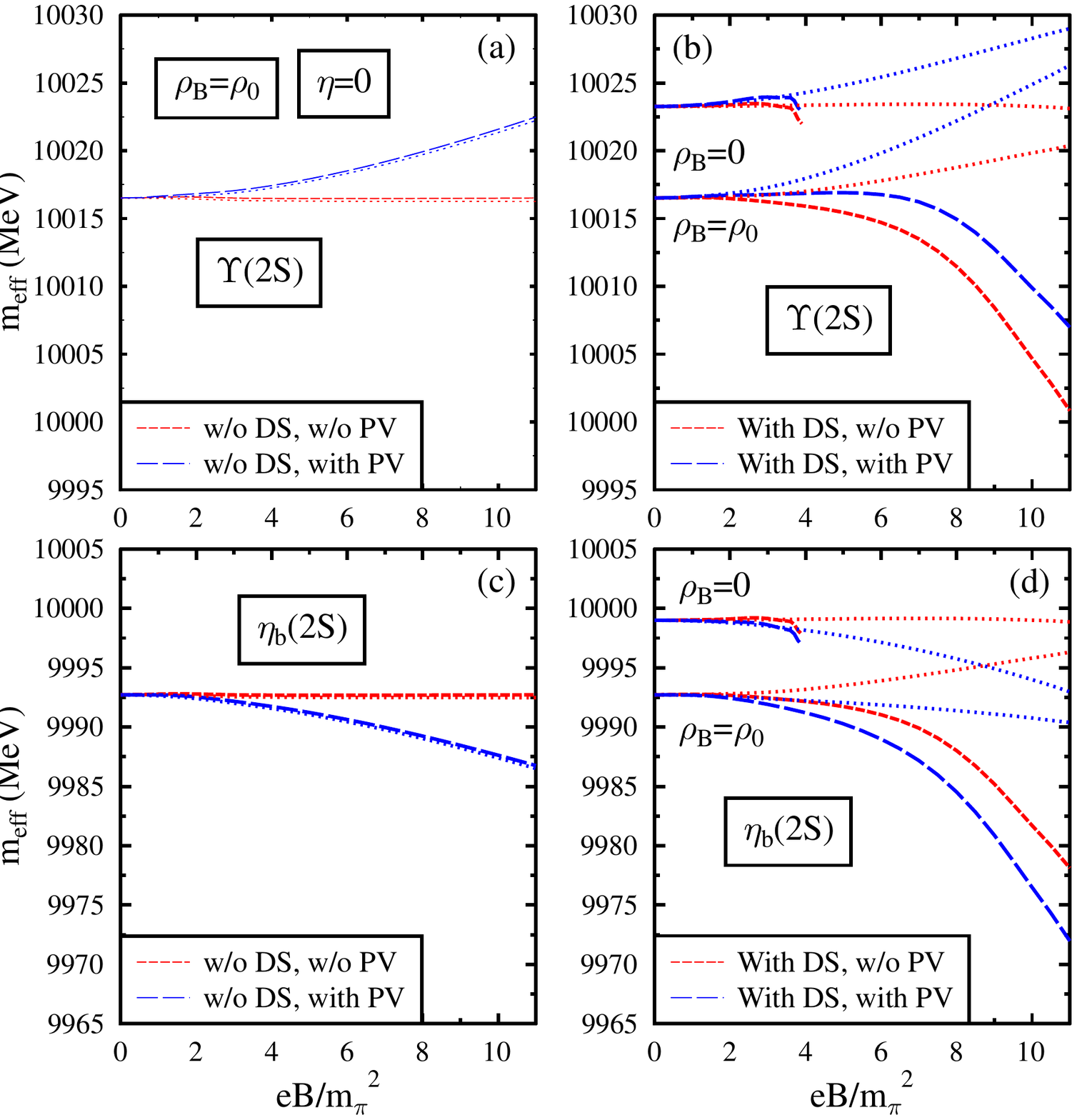}\hfill
\vskip -0.3in
    \caption{Masses (MeV) of $\Upsilon (2S)$ and $\eta_b(2S)$ are plotted as
functions of $eB/m_\pi^2$ for $\rho_B=\rho_0$ and $\eta=0$,
with and without the effects of PV mixing. 
These are shown in (a) and (c), when the Dirac sea contributions 
of nucleons (resulting in (inverse) magnetic catalysis)
are not included and in (b) and (d), while these
effects are included. In (b) and (c), the masses are also shown
for $\rho_B=0$ due to the Dirac sea conttributions.
The results are for cases when the AMMs are included,
which are compared with the cases when AMMs are not considered
(shown as dotted lines).}
    \label{mups2s_etabp_rhb0_eta0_MC}
\end{figure}

\begin{figure}
\vskip -3.2in
    \includegraphics[width=1.\textwidth]{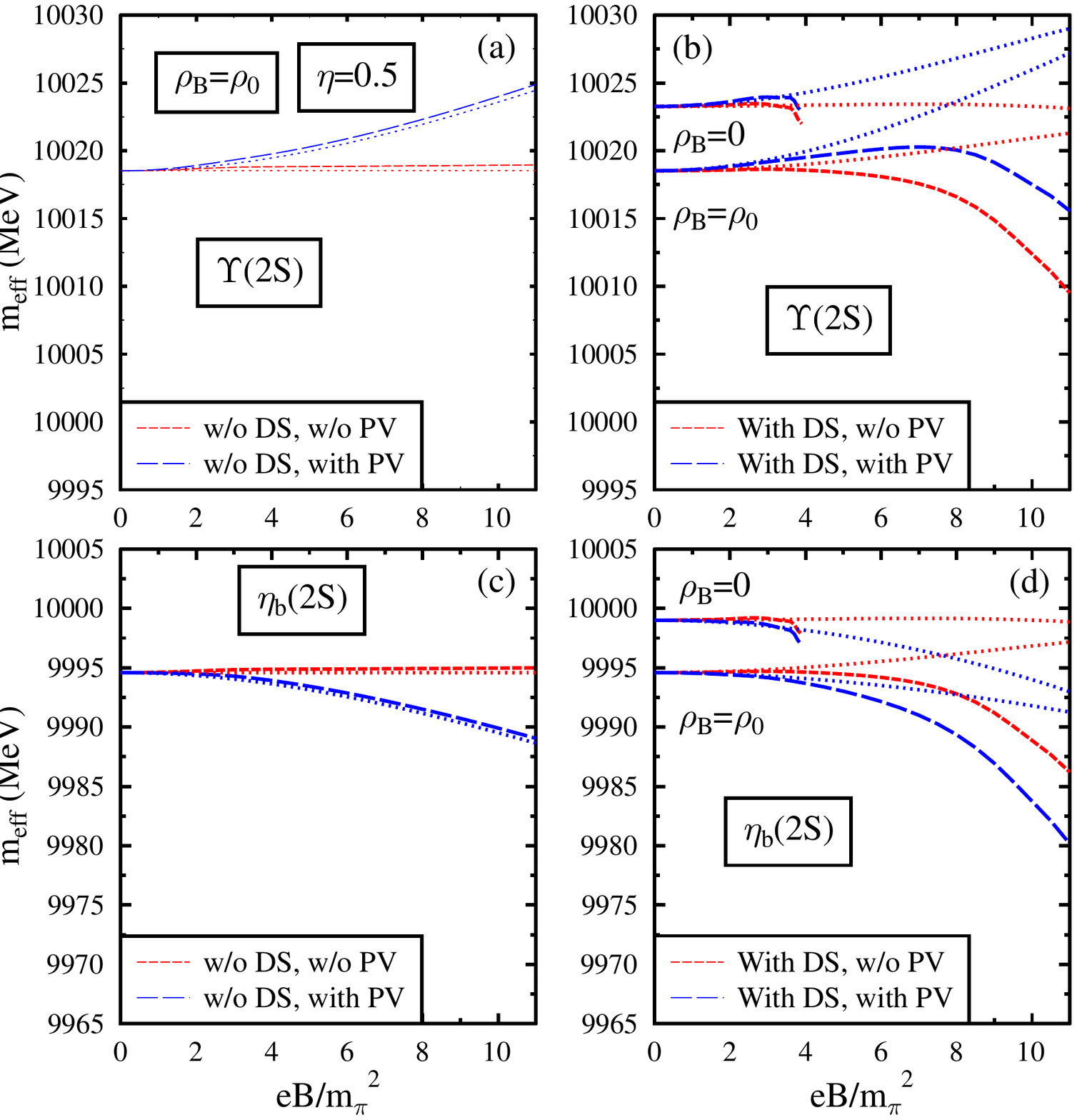}\hfill
\vskip -0.3in
    \caption{Same as Figure \ref{mups2s_etabp_rhb0_eta0_MC}, but with
$\eta$=0.5.}
    \label{mups2s_etabp_rhb0_eta5_MC}
\end{figure}

\begin{figure}
\vskip -3.2in
    \includegraphics[width=1.\textwidth]{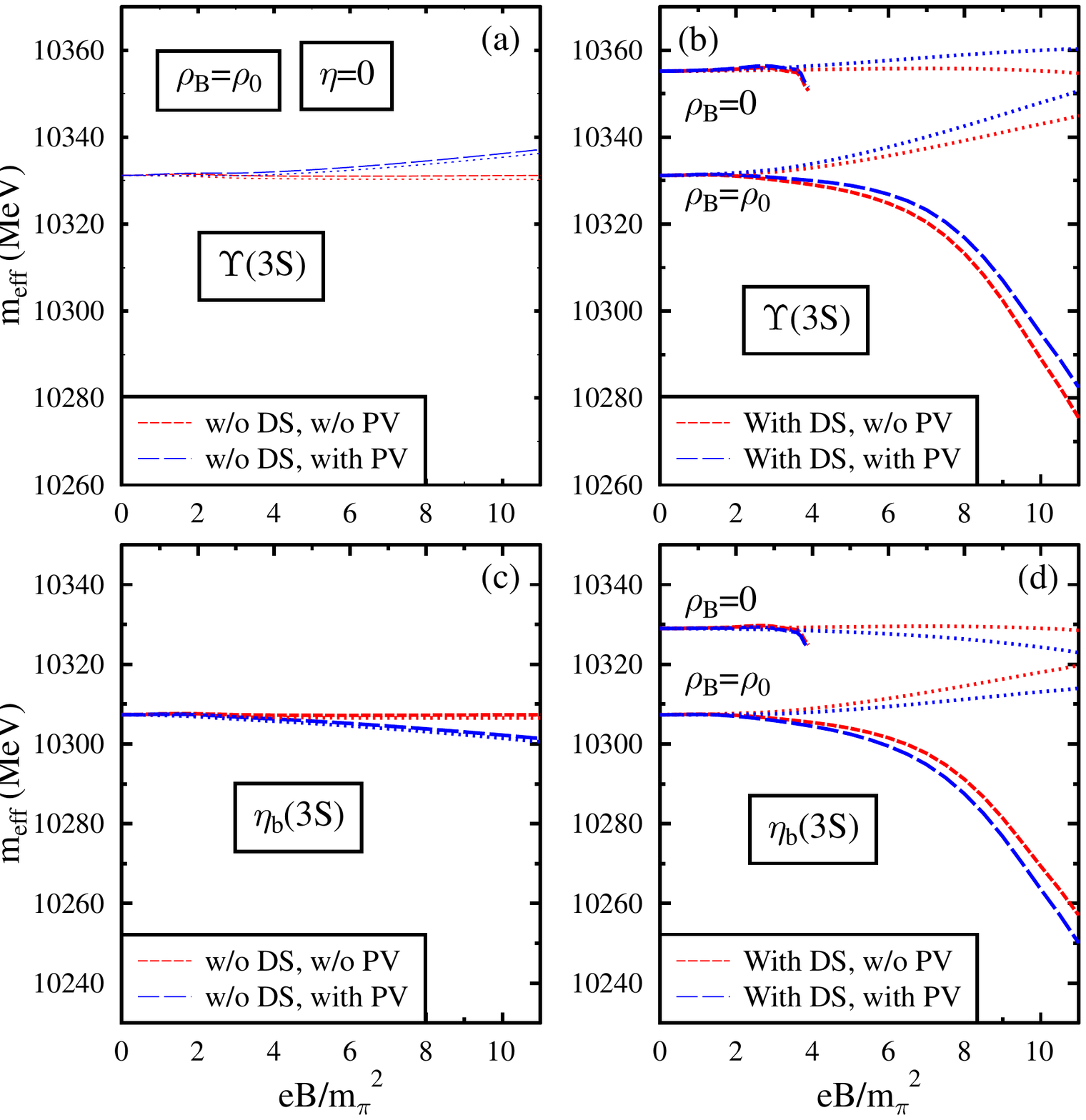}\hfill
\vskip -0.3in
    \caption{Masses (MeV) of $\Upsilon (3S)$ and $\eta_b(3S)$ are plotted as
functions of $eB/m_\pi^2$ for $\rho_B=\rho_0$ and $\eta=0$,
with and without the effects of PV mixing. 
These are shown in (a) and (c), when the Dirac sea contributions 
of nucleons (resulting in  (inverse) magnetic catalysis)
are not included and in (b) and (d), while these
effects are included. In (b) and (c), the masses are also shown
for $\rho_B=0$ due to the Dirac sea conttributions.
The results are for cases when the AMMs are included,
which are compared with the cases when AMMs are not considered
(shown as dotted lines).}
    \label{mups3s_etab3s_rhb0_eta0_MC}
\end{figure}

\begin{figure}
\vskip -3.2in
    \includegraphics[width=1.\textwidth]{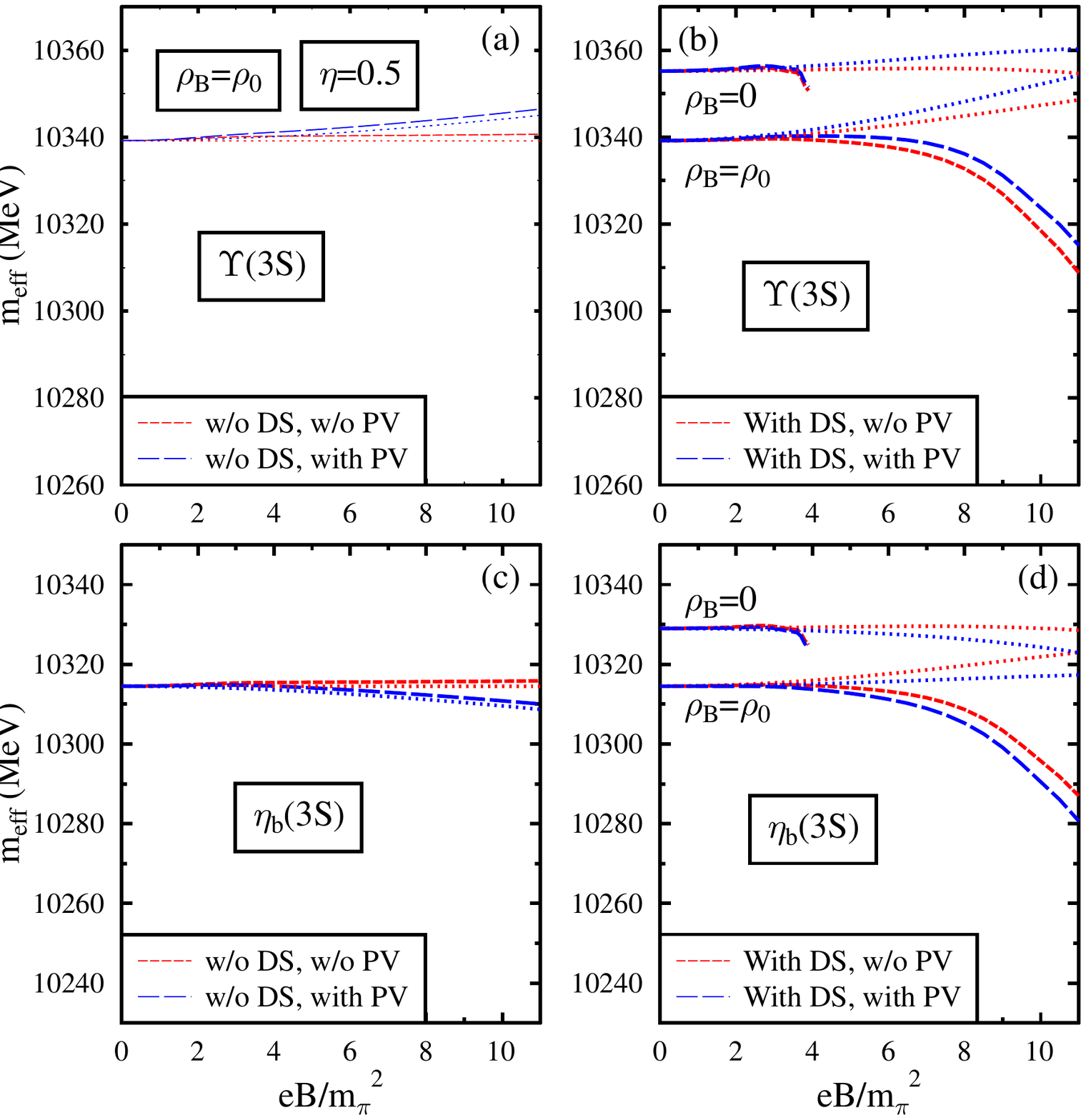}\hfill
\vskip -0.3in
    \caption{Same as Figure \ref{mups3s_etab3s_rhb0_eta0_MC}, but with
$\eta$=0.5.}
    \label{mups3s_etab3s_rhb0_eta5_MC}
\end{figure}

\begin{figure}
\vskip -3.2in
    \includegraphics[width=1.\textwidth]{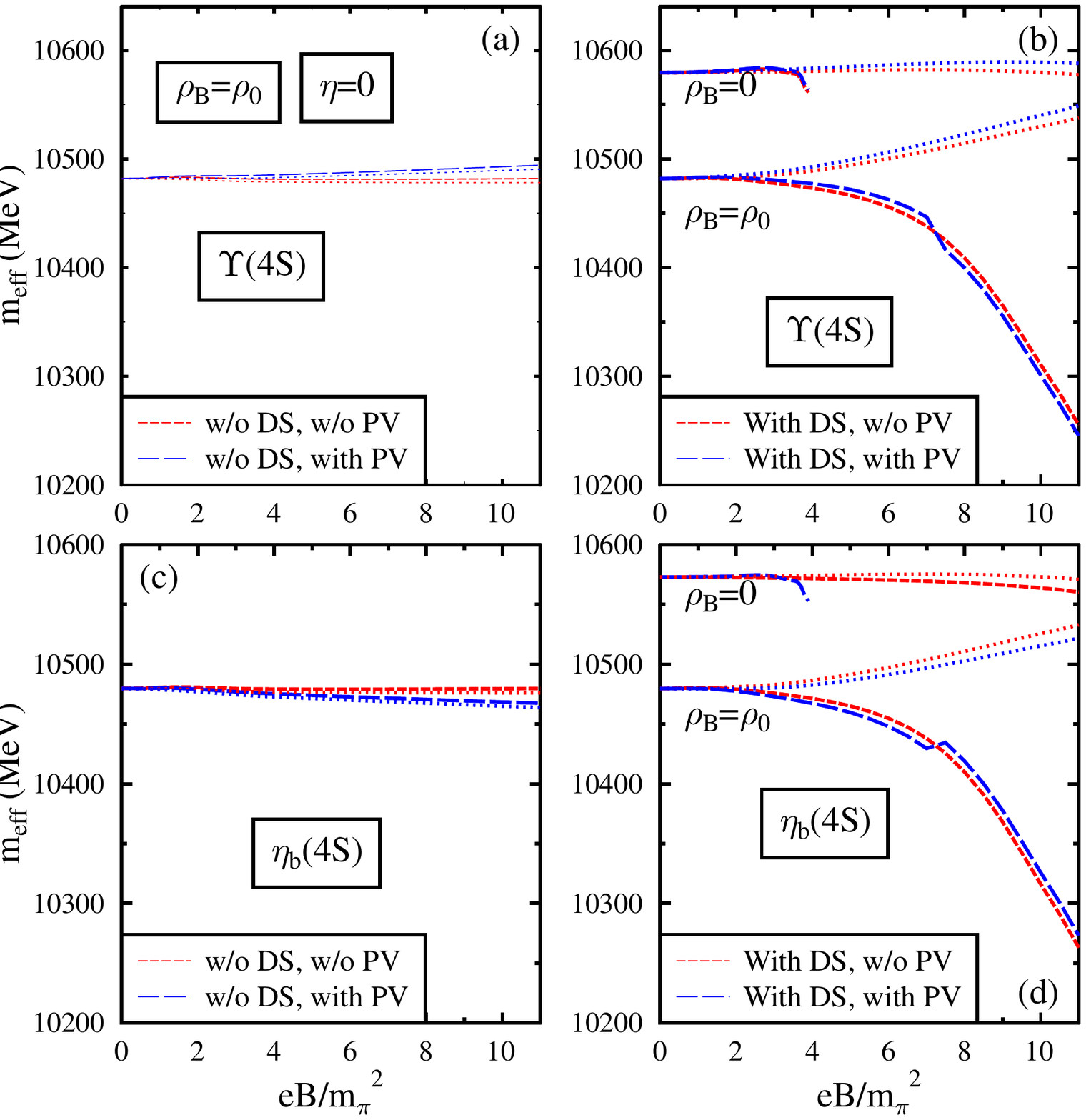}\hfill
\vskip -0.3in
    \caption{Masses (MeV) of $\Upsilon (4S)$ and $\eta_b(4S)$ are plotted as
functions of $eB/m_\pi^2$ for $\rho_B=\rho_0$ and $\eta=0$,
with and without the effects of PV mixing. 
These are shown in (a) and (c), when the Dirac sea contributions 
of nucleons (resulting in (inverse) magnetic catalysis)
are not included and in (b) and (d), while these
effects are included. In (b) and (c), the masses are also shown
for $\rho_B=0$ due to the Dirac sea conttributions.
The results are for cases when the AMMs are included,
which are compared with the cases when AMMs are not considered
(shown as dotted lines).}
    \label{mups4s_etab4s_rhb0_eta0_MC}
\end{figure}

\begin{figure}
\vskip -3.2in
    \includegraphics[width=1.\textwidth]{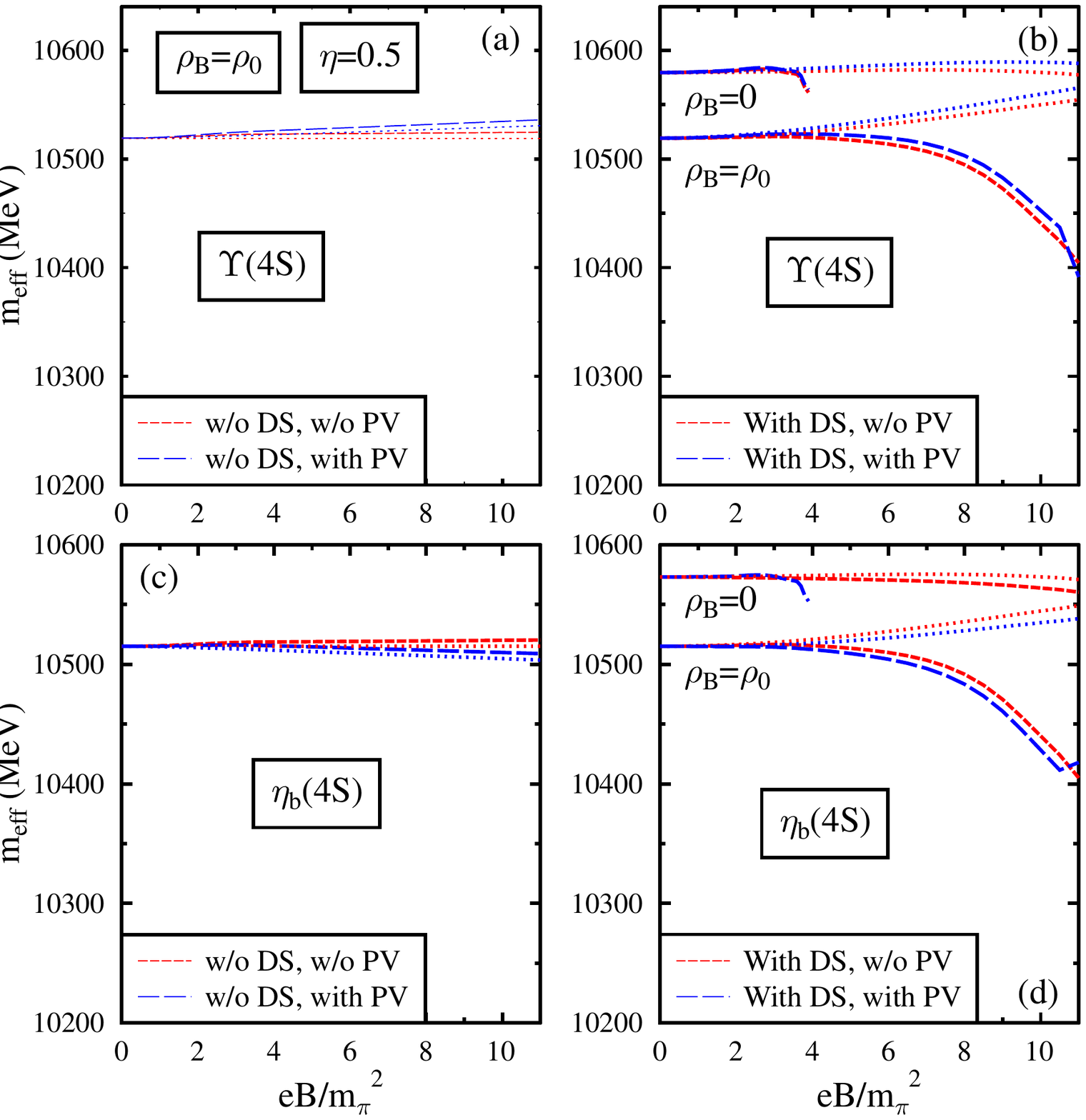}\hfill
\vskip -0.3in
    \caption{Same as Figure \ref{mups4s_etab4s_rhb0_eta0_MC}, but with
$\eta$=0.5.}
    \label{mups4s_etab4s_rhb0_eta5_MC}
\end{figure}

The mass shift of the heavy quarkonium state
is proportional to the change in the gluon condensate
in the medium. 
This is the leading order result of a study of the
heavy quarkonium state in a gluon field, assuming  
the distance between the heavy quark and antiquark
(bound by a Coulomb potential)
to be small as compared to the scale of gluonic 
fluctuations \cite{pes1,pes2,voloshin}.
The chiral effective model as used in the present work,
is based on a non-linear realization of chiral symmetry.
The model also incorporates the broken scale invariance of
QCD through a scalar dilaton field.
The dilaton field $\chi$ of 
the scale breaking term ${\cal L}_{scalebreak}$ in the chiral
effective model is related to the scalar gluon condensate
of QCD and this relation is obtained by equating the trace
of the energy momentum tensor in the chiral effective model
and in QCD \cite{amarvdmesonTprc,amarvepja,AM_DP_upsilon}.
The mass shift of the charmonium (bottomonium) 
state in the magnetized nuclear matter is hence computed
from the medium change of the dilaton field from vacuum value,
calculated within the chiral effective model,
and is given as
\cite{amarvdmesonTprc,amarvepja,AM_DP_upsilon}
\begin{equation}
\Delta m= \frac{4}{81} (1 - d) \int d |{\bf k}|^{2}
\langle \vert \frac{\partial \psi (\bf k)}{\partial {\bf k}}
\vert^{2} \rangle
\frac{|{\bf k}|}{|{\bf k}|^{2} / m_{c(b)} + \epsilon}
 \left( \chi^{4} - {\chi_0}^{4}\right),
\label{mass_shift}
\end{equation}
where
\begin{equation}
\langle \vert \frac{\partial \psi (\bf k)}{\partial {\bf k}}
\vert^{2} \rangle
=\frac {1}{4\pi}\int
\vert \frac{\partial \psi (\bf k)}{\partial {\bf k}} \vert^{2}
d\Omega.
\end{equation}
In equation (\ref{mass_shift}), $d$ is a parameter introduced
in the scale breaking term in the Lagrangian,
%%%%%%%%%added below 2nd comment of referee
 $\chi$ and $\chi_0$
are the values of the dilaton field in the magnetized medium 
and in vacuum respectively.
%%%%%%%%%added above 2nd comment of referee
The wave functions of the charmonium (bottomonium) states,
$\psi(\bf k)$ are assumed to be harmonic oscillator
wave functions, $m_{c(b)}$ is the mass of the charm (bottom) 
quark, $\epsilon=2m_{c(b)}-m$
is the binding energy of the charmonium (bottomonium) 
state of mass, $m$.
The mass shifts of the heavy quarkonium states are thus 
obtained from the values of the dilaton field, $\chi$, 
(using equation (\ref{mass_shift}). For given values of 
the baryon density, $\rho_B$, the isospin asymmetry parameter,
$\eta=(\rho_n-\rho_p)/(2\rho_B)$ (with $\rho_n$ and $\rho_p$
as the neutron and proton number densities), the magnetic field, $B$
(chosen to be along z-direction), the values of the fields 
$\chi$, $\sigma$, $\zeta$ and $\delta$ are solved from their
coupled equations of motion. The AMMs of the nucleons
are considerd in the present study 
\cite{dmeson_mag,bmeson_mag,charmonium_mag}.
In the `no sea' approximation, for the matter part of the densities 
and scalar densities of the nucleons, there are contributions 
of the Landau levels for the charged baryon, i.e., proton,
whereas the neutron interacts with the magnetic field, due to
its anomalous magnetic moment.
In the present study, the contributions of the Dirac sea are
also taken into account for the self enegies of
the nucleons, including the effects of AMMs 
of the nucleons.

\subsection{Pseudoscalar meson-Vector meson (PV) mixing}
In the presence of a magnetic field, there is mixing between
the pseudoscalar meson and vector mesons, which modifies
the masses of these mesons 
\cite{charmonium_mag_QSR,charmonium_mag_lee,Suzuki_Lee_2017,Alford_Strickland_2013,charmdw_mag,open_charm_mag_AM_SPM,strange_AM_SPM,Quarkonia_B_Iwasaki_Oka_Suzuki}.
The PV mixing leads to a drop (rise) in the mass of the
pseudoscalar (longitudinal component of the vector meson).
The mass modifications have been studied 
using an effective Lagrangian density of the form 
\cite{charmonium_mag_lee,Quarkonia_B_Iwasaki_Oka_Suzuki} 
\begin{equation}
{\cal L}_{PV\gamma}=\frac{g_{PV}}{m_{av}} e {\tilde F}_{\mu \nu}
(\partial ^\mu P) V^\nu,
\label{PVgamma}
\end{equation}
for the heavy quarkonia \cite{charmonium_mag_lee,charmdw_mag},
the open charm mesons \cite{open_charm_mag_AM_SPM}
and strange ($K$ and $\bar K$) mesons
\cite{strange_AM_SPM}.
In equation (\ref{PVgamma}), $m_{av}=(m_V+m_P)/2$, 
$m_P$ and $m_V$ are the masses 
for the pseudoscalar and vector charmonium states,
${\tilde F}_{\mu \nu}$ is the dual electromagnetic field.
In equation (\ref{PVgamma}), the coupling parameter $g_{PV}$
is fitted from the observed value of the radiative decay width, 
$\Gamma(V\rightarrow P +\gamma)$ given as
\begin{equation}
\Gamma (V\rightarrow P \gamma)
=\frac{e^2}{12}\frac{g_{PV}^2 {p_{cm}}^3}{\pi m_{av}^2},
\label{decay_VP}
\end{equation}
where, $p_{cm}=(m_V^2-m_P^2)/(2m_V)$
is the magnitude of the center of mass
momentum in the final state. 
The masses of the pseudoscalar and the longitudinal component
of the vector mesons including the mixing effects 
are given by
\begin{equation}
m^{2\ {(PV)}}_{P,V^{||}}=\frac{1}{2} \Bigg ( M_+^2 
+\frac{c_{PV}^2}{m_{av}^2} \mp 
\sqrt {M_-^4+\frac{2c_{PV}^2 M_+^2}{m_{av}^2} 
+\frac{c_{PV}^4}{m_{av}^4}} \Bigg),
\label{mpv_long}
\end{equation}
where $M_+^2=m_P^2+m_V^2$, $M_-^2=m_V^2-m_P^2$ and 
$c_{PV}= g_{PV} eB$. 
In Ref. \cite{charmdw_mag}, the effective Lagrangian term 
given by equation (\ref{PVgamma}) has been observed to lead to 
appreciable mass modifications of the pseudoscalar and the longitudinal
component of the vector charmonium states,
due to the PV mixings ($J/\psi-\eta_c$,
$\psi(2S)-\eta_c(2S)$ and $\psi(1D)-\eta_c(2S)$). 
This is observed to lead
to substantial modification of the partial decay width 
of $\psi(1D)\rightarrow D\bar D$ \cite{charmdw_mag}
due to $\psi(1D)-\eta_c(2S)$ mixing, as well as, due to
$D(\bar D)-D^* (\bar D^*)$ mixing effects 
\cite{open_charm_mag_AM_SPM}.
These mixing effects have been considered on the
masses calculated using the chiral effective model.
In the present work, the Dirac sea contributions 
are taken into account to compute the masses
of the heavy quarkonium states, additionally,
the PV mixing effects are considered for the 
masses of these mesons. 
The mixing parameter $g_{PV}$ is determined from the
observed decay widths of $V\rightarrow P\gamma$
for the open and hidden charm sector.
However, due to lack of data (radiative decay) 
for the bottomonium states,
we estimate the modifications to the masses
of the bottomonium pseudoscalar and vector mesons \cite{upslndw_mag}
due to mixing of these states in the presence of a magnetic field,  
using the Hamiltonian 
\cite{Alford_Strickland_2013,Quarkonia_B_Iwasaki_Oka_Suzuki}
\begin{equation}
H_{\rm {spin-mixing}}=-{\sum _{i=1}^2} 
{\mbox{\boldmath $\mu$}}_i
\cdot {\bf B},
\label{H_spin_mixing}
\end{equation}
which decribes the interaction of the magnetic 
moments of the quark (antiquark) with the external magnetic field.
In the above, 
${\mbox{\boldmath $\mu$}}_i
=g|e|{q_i} {\bf {S_i}}/(2m_i)$ 
is the magnetic moment of the $i$-th particle, $g$ is the Lande g-factor
(taken to be $2(-2)$ for the quark(antiquark)), $q_i$, $\bf {S_i}$,
$m_i$ are the electric charge (in units of the magnitude of the
electronic charge, $|e|$), spin and mass of the $i$-th particle
\cite{charmonium_mag_lee,Quarkonia_B_Iwasaki_Oka_Suzuki}.
This interaction leads to a drop (increase) of the mass of the 
pseudoscalar (longitudinal component of the vector meson) given as
\cite{Alford_Strickland_2013}
\begin{equation}
{\Delta M}^{PV}= \frac{\Delta E}{2} \Big ( (1+\Delta ^2)^{1/2}-1\Big),
\label{delm_PV}
\end{equation}
where $\Delta=2g|eB|((q_1/m_1)-(q_2/m_2))/\Delta E$,
$\Delta E=m_V-m_P$ is the difference in the masses 
of the pseudoscalar and vector mesons.
Here, the masses $m_V$ and $m_P$ refer to those calculated
from the medium change of the dilaton field
within the chiral effective model, using equation 
(\ref{mass_shift}). 
These masses are calculated considering the Dirac sea 
contributions. We study the PV mixing effects and consequently,
the modificaitons to the masses of the pseudoscalar
and the longitudinal component of the vector bottomonium states,
arising due to the $\Upsilon(1S)^{||}-\eta_b$, 
$\Upsilon(2S)^{||}-\eta_b(2S)$, $\Upsilon(3S)^{||}-\eta_b(3S)$,
and $\Upsilon(4S)^{||}-\eta_b(4S)$ mixing effects.

\section{Results and Discussions}
We discuss the results obtained due to the effects of 
Dirac sea contributions for the nucleons, as well as, 
PV mixing on the masses of the charmonium and bottomonium states
in magnetized isospin asymmetric nuclear matter
including the effects of the Dirac sea of nucleons.
These are obtained, using equation (\ref{mass_shift}),
from the values of the dilaton field
which are solved from the coupled equations
of motion of the scalar fields ($\sigma$, $\zeta$ and $\delta$)
and the dilaton field within the chiral effective model, 
for given values of the baryon density, $\rho_B$, isospin
asymmetry parameter, $\eta$ and the magnetic field, $B$.
The heavy quarkonium masses are studied considering 
the AMMs of the nucleons
and compared to the cases when AMMs are not taken into
account. There is observed to be enhancement of the quark condensates
(through scalar fields $\sigma$ and $\zeta$) due to Dirac sea
contributions for zero density and at $\rho_B=\rho_0$,
when the AMMs are not taken into account.  The trend
persists, when the AMMs are considered, for $\rho_B=0$, 
whereas, for $\rho_B=\rho_0$, there is observed to be the 
opposite trend of inverse magnetic catalysis.
The solution of $\chi$ is observed to be a rise (drop)
when the AMMs are neglected (included) for $\rho_B=\rho_0$.
This leads to a rise (drop) of the heavy quarkonium
masses when magnetic field, without (with) the AMMs.

In figures \ref{mjpsi_etac_rhb0_eta0_MC} and 
\ref{mjpsi_etac_rhb0_eta5_MC}, the masses of $J/\psi$ and $\eta_c$ 
are plotted for isospin symmetric nuclear matter ($\eta$=0)
and asymmetric nuclear matter (with $\eta$=0.5)
with and without PV ($J/\psi-\eta_c$) mixing effects.
The masses are plotted considering the AMMs of the nucleons
and compared with the cases when the AMMs are not taken into
account (shown as dotted lines).
These are shown in (a) and (c), when the Dirac sea contributions
are not considered. In the absence of PV mixing, there is
observed to be almost no change in the masses of $J/\psi$
and $\eta_c$ mesons. The PV mixing is observed
to lead to substantial rise (drop) in the mass
of $J/\psi^{||}(\eta_c)$ meson. The effects of
the Dirac sea contributions are shown in (b) and (d) respectively. 
There is observed to be an increase (drop) in the mass of $J/\psi$ 
with increase in the magnetic field,
without (with) accounting for the AMMs at $\rho_B=\rho_0$,
when the PV mixing is not taken into account. 
However, with PV mixing, there is an increase in the 
mass of $J/\psi^{||}$, which is observed to lead
to a rise of the mass, when the the Dirac sea,
as well as, PV mixing are taken into account,
however, the modification is much larger
at high magnetic field values, when AMMs are not
considered. For $\eta_c$, there is observed to be 
drop in the mass with increase in magnetic field
and the drop is larger with AMMs as compared to 
when the AMMs are ignored.
The effect due to PV mixing is observed to be much 
more dominant as compared to the contributions 
due to the Dirac sea effects for the masses
of both $J\psi$ and $\eta_c$ mesons. 
In (b) and (c), the results for 
masses are compared with the case
of zero baryon density. 
The plots for $\eta=0.5$ shown in figure
\ref{mjpsi_etac_rhb0_eta5_MC}, show that 
the masses of $J/\psi$ and $\eta_c$ have very small 
depndence on the isospin asymmetry of the nuclear matter.

Figures \ref{mpsip_etacp_rhb0_eta0_MC} and  
\ref{mpsip_etacp_rhb0_eta5_MC}  
show the plots of masses of $\psi(2S)$ and $\eta_c(2S)$
with and without the PV mixing effects, for $\eta=0$ and 
$\eta=0.5$ respectively. The contribution from 
(inverse) magnetic catalysis is observed to lead to appreciable
(drop) increase in the masses of these mesons.
The modifications due to the AMMs are observed 
to be quite significant. The
PV effect leads to a rise (drop) in the mass of $\psi(2S)^{||}
(\eta_c(2S))$ meson.
The effect of Dirac sea contributions is 
observed to dominate over the PV mixing contributions.
The modifications due to the isospin asymmetry
of the medium are observed to be marginal as compared
to the effects of (inverse) magnetic catalysis and PV mixing.

In figures \ref{mpsi1d_etacp1d_rhb0_eta0_MC} and  
\ref{mpsi1d_etacp1d_rhb0_eta5_MC}, we plot the 
masses of $\psi(1D)$ and $\eta_c(2S)$ for $\eta=0$
and $\eta=0.5$ for $\rho_B=\rho_0$. 
The mixing effect for $\psi(1D)-\eta_c(2S)$,
along with the Dirac sea contributions,
is observed to be lead to significant
modifications to the mass of $\psi(1D)$,
which should have observable effects on the 
production of the charmonium states
and open charm mesons, due to modification
of the decaay width of $\psi(1D)\rightarrow D\bar D$.

In figures \ref{mups1s_etab_rhb0_eta0_MC} and
\ref{mups1s_etab_rhb0_eta5_MC}, the effects of the PV mixing and
Dirac sea on the masses of the ground states
$\Upsilon (1S)$ and $\eta_b$ are shown at $\rho_B=\rho_0$
and for $\eta=0$ and $\eta=0.5$ respectively.
There is observed to be a rise (drop) in the mass
of $\Upsilon (1S)^{||}$($\eta_b$), when the effects of
DS and PV mixing are both considered, and, the AMMs
of nucleons are taken into account. When only
the DS effects are considered, the masses of
$\Upsilon(1S)$ and $\eta_b$ are observed 
to drop (rise) with increase in magnetic field,
with (without) AMMs of the nucleons. 
For the excited bottomonium states, there is observed
to be drop in masses of both the pseudoscalar and
the vector mesons, when only DS effects are taken 
into account. The effect  of isospin
asymmetry is observed to lead to smaller
modifications of these excited  states,
as might be seen from the figures 
\ref{mups2s_etabp_rhb0_eta0_MC}, \ref{mups3s_etab3s_rhb0_eta0_MC}
and  \ref{mups4s_etab4s_rhb0_eta0_MC} for $\eta=0$,
and, from the figures 
\ref{mups2s_etabp_rhb0_eta5_MC}, \ref{mups3s_etab3s_rhb0_eta5_MC}
and  \ref{mups4s_etab4s_rhb0_eta5_MC} for $\eta=0.5$.
The effect of PV mixing leads to a positive
(negative) contribution to the mass of $\Upsilon(NS)^{||}$ ($\eta_b(NS)$,
$N=1,2,3,4$, with magnetic field, the mass shifts
found from the equation \ref{delm_PV}.
However, the PV mixing is observed to have the opposite
trend for the $\Upsilon (4S)-\eta_c(4S)$ mixing 
at higher values of magnetic field
($eB > 7 m_\pi^2$) for $\rho_B=\rho_0$
in symmetric nuclear matter ($\eta$=0), as can be
seen from figure \ref{mups4s_etab4s_rhb0_eta0_MC}.
This is because the effective mass of the $\Upsilon (4S)$
turns out to be smaller than that of $\eta_b(4S)$,
as calculated within the chiral effective model,
which makes $\Delta E$ (and hence $\Delta M^{PV}$)
of equation (\ref{delm_PV}) to be negative.
However, it is observed that the PV mixing has 
much smaller contribution as compared to the Dirac sea
contributions for the excited bottomonium states,
contrary to the charm sector, where both PV mixing and the
DS contributions are observed to be important. 
The (inverse) magnetic catalysis, along with
PV mixing can thus modify the heavy quarkonium
decay widths to the open heavy flavour mesons,
and hence can affect the production of the charm
and bottom mesons in non-central ultra-relativistic
heavy ion collision experiments, due to existence
of strong magnetic fields.

\section{summary}
To summarize, we have studied the masses of the heavy quarkonium
states in magnetized nuclear matter. The masses are calculated
within a chiral effective model from the medium change 
of a scalar dilaton field, which mimics the gluon condensates
of QCD. The effects if Dirac sea of nucleons are taken into
consideration, which lead to increase (decrease) of the magnitudes
of the scalar fields (thus leading to increase
in the light quark condensates), with rise
in the magnetic field, an effect called (inverse) magnetic 
catalysis. For zero density, one oberves the magnetic catalysis,
when the anomalous magnetic moments (AMMs)
are (not) considered. However, at $\rho_B=\rho_0$,
there is observed to be (inverse) magnetic catalysis,
when the AMMs are (included) ignored. 
In the presence of the magnetic field, there
are further modifications to the masses of the chamronium and 
bottomonium states due to PV mixing. The effects from 
isospin asymmetry on the in-medium
masses of heavy quarkonia are observed to be small
and the dominant contributions due to magnetic field effects
are observed to be arising from the effects of
PV mixing as well as Dirac sea conttributions.
These should have observable consequences
on the production of heavy quarkonium states 
and open heavy flavour mesons, as these are created 
at the early stage of the non-central ultra-relativistic 
heavy ion collision experiments, when the magnetic field
can be large.

\end{document}